
\let\mycases=\cases
\input vanilla.sty
\font\tenbf=cmbx10
\font\tenrm=cmr10
\font\tenit=cmti10

\font\ninerm=cmr9

\font\eightrm=cmr8
\font\eightit=cmti8
\font\sevenrm=cmr7
\TagsOnRight
\hsize=5.0truein
\vsize=7.7truein
\parindent=15pt
\nopagenumbers
\baselineskip=10pt
\centerline{\tenbf RECENT RESULTS OF MULTIMAGNETICAL SIMULATIONS}
\centerline{\tenbf OF THE ISING MODEL\footnote"$*$"
{\eightrm\baselineskip=10pt To appear in the proceedings of the Workshop
on ~~Dynamics of First Order Phase Transitions'',J{\"u}lich, June 1-3,
1992, to be published in Int. J. Mod. Phys. C}}
\vglue 24pt
\centerline{\eightrm U. HANSMANN, B.A. BERG}
\baselineskip=12pt
\centerline{\eightit
 Department of Physics and  Supercomputer Computations Research Institute}
\baselineskip=10pt
\centerline{\eightit
 The Florida State University, Tallahassee,  FL 32306, USA}
\vglue 10pt
\centerline{\eightrm and}
\vglue 10pt
\centerline{\eightrm T. NEUHAUS}
\baselineskip=12pt
\centerline{\eightit
 Fakult\"at f\"ur Physik, Universit\"at Bielefeld}
\baselineskip=10pt
\centerline{\eightit D-W-4800 Bielefeld, FRG}
\vglue 16pt
\centerline{\eightrm ABSTRACT}
{\rightskip=1.5pc
 \leftskip=1.5pc
 \eightrm\baselineskip=10pt\noindent
To investigate order-order interfaces, we perform multimagnetical Monte
Carlo simulations of the $2D$ and $3D$ Ising model.
Stringent
tests of the numerical methods are performed by reproducing with high
precision exact $2D$ results. In the physically more interesting
$3D$ case we estimate the amplitude $F^s_0$ of the critical interfacial
tension.
\vglue 5pt
\noindent
{\eightit Keywords}\/: Monte Carlo Simulations; Ising model;
                       Interfaces; Surface tension.
\vglue 12pt}
\baselineskip=13pt
\line{\tenbf 1. Introduction \hfill}
\vglue 5pt
Since long,  there has been
continuous interest in the properties of interfaces in
Ising models. Past numerical studies were, however, hampered by a problem
of principle. The surface tension per unit area $F^s$ between different
states has a finite value. Thus in the canonical ensemble, where  one
samples with the Boltzmann weights ${\cal{P}}^B \propto
e^{-\beta H}$, configurations containing interfaces with an area $A$ are
supressed by exponentiall factors $e^{-AF^s}$. Corresponding there is an
exponentially fast increase of the tunneling time between pure phases when
the system is simulated with {\it local} Monte Carlo (MC) algorithms.
Recently this difficulty was overcome by
the proposal to perform the MC simulation for a multimagnetical ensemble,$^1$
 a natural extension of the multicanonical ensemble
introduced in (Ref.~2). The original canonical ensemble
is then obtained by re-weighting,$^{3,4}$
while the slowing down becomes reduced to a power law close to $\sim V^2$.

For our models, spins $s_i = \pm 1$ are defined on  sites
of a square lattice of volume $V = L^D$ with periodic boundary conditions
and the symbol $<i,j>$ is used to denote nearest neighbors. The partition
function of the Ising model is given by
$$
Z\ =\ \sum_{\text{configurations}} \exp ( - \beta H) . \tag 1.1
$$
The Ising Hamiltonian $H$
$$
H\ =\ H_I\ -\ h M ~~~\text{ with}~~~ H_I = - \sum_{<i,j>} s_i s_j
{}~~\text{ and}~~ M = \sum_i s_i \tag 1.2
$$
contains the nearest neighbor interaction term $H_I$ and a term
which couples the external magnetic field $h$ to the magnetization.
In our case is $h = 0$.
In the broken region $\beta > \beta_c$ the magnetic probability
densities $P_L (M)$ are double peaked and  $P_L (M) = P_L (-M)$, as the model
is globally $Z(2)$ symmetric.  The positions of
the maxima are $\pm M^{\max}_L$ and the
distribution takes its minimum at $P_L^{\min} = P_L (0)$.

The surface tension $F^s$ is the free energy per unit
area of the interface between coexisting phases. Here we consider
order-order interfaces in the broken region $\beta > \beta_c$.
The interface tension $F^s$ can be defined from the finite size scaling
(FSS) of the ratio $P_L^{\min} / P_L^{\max}$, which for large $L$ takes
the form$^{5,6}$
$$
{P^{\min}_L \over P^{\max}_L}\ =\
\text{const}\ L^p\ \exp \left[ -2 L^{D-1} F^s (1+ O(L^{-1})) \right]
\tag 1.3
$$
on lattices with periodic boundary conditions. Physically this
definition assumes that at values of the mean magnetization $M = 0$ a
rectangular domain, enclosed by two interfaces and spanning the lattice
via the periodic boundary condition, is formed.
The finite lattice interfacial tension is then defined as
$$
F^s_L\ =\ {1\over 2 L^{D-1}}\ln {P^{\min}_L \over P^{\max}_L} , \tag 1.4
$$
and $F^s = \lim_{L\to\infty} F^s_L$.
So the final estimate of $F^s$ requires a FSS extrapolation towards
$L=\infty$. From equation ``(1.3)'' one gets the fit:
$$
F^s_L\ =\ F^s + {a \over L^{D-1}} + {b\ \ln (L) \over L^{D-1}} . \tag
1.5
$$
\vglue 12pt
\line{\tenbf 2. Multimagnetical Monte Carlo\hfill}
\vglue 5pt
To get a representative sample of interfaces
one has to generate many configurations with $M=0$ on large lattices.
But, in a canonical simulation this is almost impossible.
As an example see figure 1. The logarithmic
scale displays that more than twenty orders of magnitude are involved, i.\
e.\ $P^{\min}_L/ P^{\max}_L < 10^{-20}$ for the largest lattice with
$L =100$. The standard MC algorithm would only sample configurations
corresponding to $P^{\min}_{100}$ if one could generate of the order $O
(10^{20})$ or more statistical independent configurations.
\hfill \break
\midinsert\vskip 8cm
\centerline{\eightrm Fig. 1.\,\,\, Boltzmann probability distribution $P_L (M)$
 of the magnetization for different lattice sizes.}
\endinsert

 Here we overcome this
difficulty by sampling configurations with a multimagnetical weight factor
$$
{\cal P}^{mm}_L (M)\ = \exp ( \alpha_L (M) + h_L (M) \beta M - \beta H_I )
 ~~~\text{for}~~~ -V \le M \le V \tag 2.1
$$
instead of sampling with the Boltzmann factor ${\cal P}^B = \exp (-\beta H_I)$.
The function $h_L (M)$ defines an $M$-dependent effective external magnetic
field such that the resulting multimagnetical probability density is flat
for $-M_L^{\max}\le M\le M_L^{\max}$.
We chose the ansatz:
$$
h_L (M)\ =\ \mycases{0 ~~\text{for}~ M\le -M^{\max}_L ~\text{and}~ M\ge
M^{\max} ;
\cr
2^{-1}{\beta}^{-1} \left[ \log P_L (M) - \log P_L (M+2) \right]
                                     ~\text{elsewhere}. } \tag 2.2
$$
The function $\alpha_L (M)$ is then defined by
the recursion relation
$$
\alpha_L (M+2)\ =\  \alpha_L (M) + \beta \left[ h_L(M)-h_L(M+2) \right]\
(M+2),
\ \alpha_L (-M_L^{\max}) = 0 . \tag 2.3
$$
Once $h_L (M)$ is given, $\alpha_L (M)$ follows automatically.
With this choice the resulting multimagnetical
probability density will be approximately flat:
$$
P_L^{mm} (M)\ =\ n(M)\ {\cal P}_L^{mm} (M)\ \approx\ \text{const}. \tag
2.4
$$
Here $n(M)$ is the magnetical density of of states (for fixed temperature
$\beta^{-1}$). The standard Markov process is well-suited to generate
configurations which are in equilibrium with respect to this
multimagnetical distribution. The canonical probability density
$P_L (M)$ is obtained from $P_L^{mm} (M)$ by re-weighting$^{3,4}$:
$$
P_L (M)\ =\ c\ P_L^{mm} (M) \exp ( - \alpha_L (M) - h_L (M) \beta M ).
\tag 2.5
$$
The constant $c$ is obtained by imposing the appropriate
normalization on $P_L (M)$.
\midinsert\vskip 8cm
{\eightrm Fig. 2.\,\,\,
Tunneling times versus lattice size. The upper data are for the
standard MC and the broken line extrapolates by means of a fit into the
region where no data exist. The lower data points are obtained with our
multimagnetical MC.}
\endinsert
 For small systems
$P_L (M)$ can  be calculated by performing standard MC
simulations, and $h_L (M)$ follows directly from  ``Eq. 2.2''.
On larger systems we get $h_L (M)$ by making
every time a FSS prediction of $ P_L (M)$ from the already controlled
smaller systems.
\hfill \break

To compare the efficiency of our method with standard MC we measured for
the $2D$ Ising model with $\beta = 0.5$ the tunneling time $\tau^t_L$.
As before$^{1,2}$ we define the tunneling time $\tau^t_L$ as
the average number of updates needed to get from a configuration
with $M=M^{\min}=-M^{\max}$ to a configuration with $M = M_L^{\max}$
and back. In figure 2 we display on a log-log scale both the tunneling
times for the multimagnetical MC and the heatbath algorithm. While there is
an exponentiall fast increase of $\tau^t_L$ for the heatbath algorithm, the
increase of $\tau^t_L$ is for the multimagnetic MC of the type of a power
law divergence. The ratio
$$
R= \tau^t_L (\text{heatbath})/ \tau^t_L (\text{multicanonical}) \tag 2.6
$$
is a direct measure for the improvement due to our method. $R$ increases
from a factor 4 for the smallest lattice ($L=2$) up to $R \approx 450$ for
$L =16$, the largest lattice size  where it was with our statistics possible to
get data from standard
MC. The extrapolation to $L=100$ yields $R \approx 6.1 \times 10^{15}$,
i.e. an improvement by more than fifteen orders of magnitude.

\vglue 12pt
\line{\tenbf 3. Numerical Results \hfill}
\vglue 5pt
\midinsert\vskip 8cm
\centerline{\eightrm Fig. 3.\,\,\,
$2D$ effective tensions.  The lines are our fits}
\endinsert
For the two dimensional Ising model we performed multimagnetical
simulations at the critical temperature
$\beta = {\beta}_c = \ln (1+\sqrt{2})/2 = 0.44068...$, at $\beta = 0.47$
and 0.5 with at least $4 \times 10^6$ sweeps per lattice size.
 In each run additional $200,000$ initial sweeps without measurements
were performed for reaching equilibrium with respect to the multimagnetical
distribution. We compared our $L=\infty$ estimates of the tension with
the exact values which follow
from Onsager's equation$^7$
$$
F^s\ =\  2\beta - \ln \left[ {1+e^{-2\beta} \over 1-e^{-2\beta}} \right],
{}~~~(\beta \ge \beta_c) .
\tag 3.1
$$
Both are collected in table 1.
Good agreement is found in all cases.

\smallskip
\centerline{\eightrm Table 1. Surface tensions $F^s$ for the $2D$ Ising
 model}
$$\vbox{\eightrm\halign{\hfil#\quad\hfil &#\hfil\quad
 &\hfil#\hfil\cr
\multispan3\hrulefill\cr
\hfil$\beta$\hfil &\hfil$ F^s_{\text{mm}}$\hfil &\hfil $F^s_{\text{exact}}$
\hfil \cr
\multispan3\hrulefill\cr
$\beta_{c}$ & 0.00033(34) & 0 \hfil \cr
\hfil 0.47  \hfil& 0.11526(80) & 0.11492... \hfil \cr
\hfil 0.5 \hfil & 0.2281(5)\hfil & 0.22806... \hfil \cr
\multispan3\hrulefill\cr}}$$
\medskip\noindent

\midinsert\vskip 8cm
\centerline{\eightrm Fig. 4.\,\,\,
$3D$ effective tensions.  The lines are our fits.}
\endinsert
In the case of the $3D$ Ising model
we performed simulations at $\beta = 0.227$, 0.232 and 0.2439 with at least
$4 \times 10^6$ sweeps for every lattice size.
As in the $2D$ case 200,000
additional, initial sweeps were performed in each run for reaching
equilibrium.
In figure 4 we display the effective tensions as functions of the
lattice size together with asymptotic fits. We notice that finite size
effects play a more important role in three dimensions than two and are
more complicated, too. The non-monotone behaviour shows that it is
necessary to use large enough lattices to estimate the interface tension.

The value $\beta = 0.232$ was chosen, because it enables an comparison
with the recent literature. In (Ref.~8) cluster improved estimators
were used to calculate correlations for $L=8-14$ in a cylindrical geometry.
Fitting the obtained tunneling mass gaps yields $F^s = 0.03034 \pm 0.00015$
for the surface tension, when systematic errors are admitted,
this is consistent with our value.
 All our $F^s$ (see table 2) values are much higher that the old estimates
of (Ref.~6), which had to rely on far too small lattices. Our estimates are
also
consistent with the valueses presented by M{\"u}nster$^{9}$ and
Potvin$^{10}$ in other talks on this workshop.
\smallskip
\centerline{\eightrm Table 2. Surface tensions $F^s$ and the amplitude
$F^s_0$ for the $3D$ Ising model}
$$\vbox{\eightrm\halign{\hfil#\quad\hfil &#\hfil\quad
 &\hfil#\hfil\cr
\multispan3\hrulefill\cr
\hfil $\beta$ \hfil &\hfil $F^s$\hfil &\hfil $F^s_{0}$ \hfil \cr
\multispan3\hrulefill\cr
 0.227& 0.01293(17) & 1.455(9) \hfil \cr
\hfil 0.232  \hfil& 0.03140(14) & 1.581(7) \hfil \cr
\hfil 0.2439 \hfil & 0.07403(30)\hfil & 1.513(6) \hfil \cr
\multispan3\hrulefill\cr}}$$
\medskip\noindent

Of particular interest in the $3D$ Ising model are amplitude
ratios$^{11,12}$
which involve the amplitude $F^s_0$ of the critical $3D$ interfacial
tension, because they can be compared with a variety of experimental
results$^{9,13,14}$ for fluids which are supposed to populate the
Ising model universality class.
The other amplitudes are fairly accurately known,
and the uncertainty in the ratios stems mainly from $F^s_0$.
Old estimates of $F^s_0$ (Ref.~6)
 led to amplitude ratios which are in disagreement with the
experimental results$^{13,14}$,
whereas Mon obtained consistency by calculating the excess
energy between a system with an interface imposed by an antiperiodic
boundary condition and a system without such an interface and with periodic
boundary conditions.$^{15}$

The
asymptotic behaviour of the interface tension for $\beta\to\beta_c$ is
$$
F^s\ =\ F^s_0 t^\mu  ,  \tag 3.2
$$
where $t=(1-\beta_c/\beta)$ is the reduced temperature and
 $\mu = 1.26$ in $3D$. Our $t$-values are small enough to apply this
 formula.
The results for the amplitude are included in table 2.
By averaging over our three values we get
$$
F^s_0\ =\ 1.52 \pm 0.05 , \tag 3.3
$$
compared to Mon's value $F^s_0 = 1.58 \pm 0.05$ (Ref.~15).

\vglue 12pt
\line{\tenbf 4. Conclusion\hfill}
\vglue 5pt
Multimagnetical simulations allow to study the magnetic
probability density in the broken region with a hitherto unreached
precision. Our numerical calculations for the $2D$ Ising model  agree well
with the exact results. For the $3D$ Ising model we obtain new surface
tension estimates,which agree resonable with (Ref.~8,9,10), and for our
amplitude  good agreement is found with Mon (Ref.~15).
\vglue 12pt
\line{\tenbf Acknowledgements \hfil}
\vglue 5pt
Our simulations were performed on the SCRI
cluster of fast RISC workstations. This research project was partially
supported by
the U. S. Department of Energy under contract DE-FG05-87ER40319,
DE-FC05-85ER250000 and by FSU's COFRS program.  U. Hansmann is supported
by Deutsche Forschungsgemeinschaft under contract H180411-1.
\vglue 12pt
\line{\tenbf References \hfil}
\vglue 5pt
\medskip
\ninerm
\baselineskip=11pt
\frenchspacing
\item{1.} B. Berg, U. Hansmann and T. Neuhaus,
               SCRI-91-125, submitted to Phys. Rev. B
\item{2.} B. Berg and T. Neuhaus, Phys. Lett. {\bf B267}, 249 (1991);
               Phys. Rev. Lett. {\bf  68}, 9 (1999).
\item{3.} B. Baumann and B. Berg, Phys. Lett. {\bf  164B}, 131 (1985);
          B. Baumann, Nucl. Phys. {\bf  B285}, 391 (1987).
\item{4.} A.M. Ferrenberg and R.H. Swendsen, Phys. Rev. Lett.
              {\bf  61}, 2635 (1988); {\bf 63 }, 1658(E) (1989), and
              references given in the erratum.
\item{5.} K. Binder, Z. Phys. {\bf B43}, 119 (1981).
\item{6.} K. Binder, Phys. Rev. {\bf A25}, 1699 (1982).
\item{7.} L. Onsager, Phys. Rev. {\bf 65}, 117 (1944).
\item{8.} H. Meyer-Ortmanns and T. Trappenberg, J. Stat. Phys. {\bf 58}
	  (1990), 185
\item{9.} G. M{\"u}nster, this volume
\item{10.} J. Potvin, this volume
\item{11.} S. Fisk and B. Widom, J. Chem. Phys. {\bf 50}, 3219 (1969).
\item{12.} D. Stauffer, M. Ferer and M. Wortis, Phys. Rev. Lett.
                   {\bf 29}, 345 (1972).
\item{13.} M.R. Moldover, Physical Review A {\bf 31}, 1022 (1984);
              H. Chaar, M.R. Moldover and J.W. Schmidt, J. Chem. Phys.
              {\bf 85}, 418 (1986).
\item{14.} H.L. Gielen, O.B. Verbeke and J. Thoen, J. Chem. Phys.
              {\bf 84}, 6154 (1984).
\item{15.} K.K. Mon, Phys. Rev. Lett. {\bf 60}, 2749 (1988).
\vfil\supereject
\bye